\documentclass[11pt]{article}
\usepackage{moriond,epsfig}
\usepackage{xspace}
\usepackage{url}

\def\Journal#1#2#3#4{{#1} {\bf #2}, #3 (#4)}

\def\PLB{{\em Phys. Lett.}  B}
\def\PRL{\em Phys. Rev. Lett.}
\def\PRD{{\em Phys. Rev.} D}

\def\be{\begin{equation}}
\def\ee{\end{equation}}
\def\bea{\begin{eqnarray}}
\def\eea{\end{eqnarray}}

\newcommand{\dzero}{D\O\xspace}
\newcommand{\fb}{fb$^{-1}$\xspace}

\newcommand{\ppbar}{\ensuremath{p\bar{p}}\xspace}
\newcommand{\ttbar}{\ensuremath{t\bar{t}}\xspace}

\newcommand{\pt}{\ensuremath{p_{T}}\xspace}
\newcommand{\ptmiss}{\ensuremath{p \kern-0.5em\slash_{T}}\xspace}
\newcommand{\mtop}{\ensuremath{m_{t}}\xspace}

\newcommand{\mH}{\ensuremath{m_{H}}\xspace}
\newcommand{\mW}{\ensuremath{m_{W}}\xspace}

\newcommand{\alpgen}{{\sc alpgen}\xspace}
\newcommand{\pythia}{{\sc pythia}\xspace}

\newcommand{\etal}{\emph{et al.}\xspace}
\newcommand{\MSbar}{\ensuremath{\overline{\mathrm{MS}}}\xspace}
\begin{document}
\hspace{4in} \mbox{FERMILAB-CONF-11-341-E}
\vspace*{4cm}
\title{Top Quark Mass Measurements at the Tevatron}

\author{Zhenyu Ye (for CDF and \dzero collaborations)}

\address{Fermilab, Batavia 60510, U.S.A.}

\maketitle\abstracts{
We report the latest results on the top-quark mass and on the top-antitop mass difference from the CDF and \dzero collaborations using data collected at the Fermilab Tevatron \ppbar collider at $\sqrt{s}=1.96$ TeV.
We discuss general issues in top-quark mass measurements and present new results from direct measurements and from top-pair production cross-section.
We also report new results on the top-antitop mass difference.}

\section{Introduction}
The observation of the top quark in 1995~\cite{topdiscovery} confirmed the existence of the six quarks in three generations of fermions expected in the standard model (SM) of particle interactions. 
The large mass of the top quark (\mtop), corresponding to a Yukawa coupling to the Higgs boson equal to $1$ within the current uncertainties, suggests a special role for the top quark in the breaking of electroweak symmetry.
It is therefore not surprising that the precise determination of the mass of the top quark has received great attention.  
The interest in the top-quark mass also arises from the constraint imposed on the mass of the Higgs boson (\mH) from the relationship among the values of \mtop, \mH, and the SM radiative corrections to the mass of the $W$ boson (\mW)~\cite{lepewwg}. 
We report the latest Tevatron results on the top-quark mass from direct measurements in section \ref{mass}, and those from top-pair production cross-section in section \ref{xsection}.

The SM is a local gauge-invariant quantum field theory conserving CPT invariance.
In the above measurements we assume the top and antitop quarks have the same mass since a difference in the mass of a particle and its antiparticle would constitute a violation of CPT invariance.
Because of its mass, the lifetime of the top quark is much shorter than the time-scale of hadronization.
The top quark can decay before interacting, making it the only quark whose characteristics can be studied in isolation.
Thus the top quark provides a unique opportunity to measure directly the mass difference between a quark and its antiquark.
We present results on the top-antitop mass difference ($\Delta M=M_t-M_{\bar{t}}$) in section \ref{difference}.

\section{Direct Measurements of Top-Quark Mass\label{mass}}
Direct measurements of the top-quark mass have been performed in the dilepton ($t\bar{t}\rightarrow l^+\nu_ll^-\bar{\nu}_lb\bar{b}$), lepton+jets ($t\bar{t}\rightarrow l\nu_lq\bar{q}'b\bar{b}$) and all hadronic ($t\bar{t}\rightarrow q\bar{q}'q''\bar{q}'''b\bar{b}$) channels~\footnote{Here $l=e,\,\mu$.}.
Events are selected after requiring there are 2, 1 or 0 high transverse momentum (\pt) leptons and at least 2, 4 or 6 high \pt jets in the dilepton, lepton+jets or all hadronic channel, respectively, and requiring large missing transverse momentum (\ptmiss) in the dilepton and lepton+jets channels.
A requirement on the minimum number of jets identified as $b$-quark jets is often applied in the lepton+jets and all hadronic channels.
After putting additional kinematic cuts to further suppress background contribution, the remaining background is dominated by the contribution from $Z/\gamma^*$+jets, $W$+jets, or multijet production~\cite{Silvia}.

Two approaches are most often used in direct measurements of the top-quark mass. 
One approach is the so-called matrix element (ME) method.
It is based on the likelihood to observe a sample of selected events in the detector.
The likelihood is calculated as a function of $m_t$ using theoretically predicted differential cross-sections and experimentally determined detector resolutions.
An integration is performed over all possible momentum configurations of the final state particles with transfer functions that relates an assumed final-state momentum configuration to the measured quantities in the detector.
The other approach is the template method, which is based on a comparison of Monte Carlo (MC) templates for different assumed values of $m_t$ with distributions of kinematic quantities measured in data.
In both approaches $m_t$ and its uncertainty extracted from data are corrected by calibration curves obtained from MC pseudo-experiments.
The ME measurement requires significantly much more computation time than a template measurement, but has a superior performance in terms of the measurement statistical uncertainty, and has consistently provided the single best result for the \mtop measurements.

In direct top-quark mass measurements, the dominant contribution to the total uncertainty comes from the uncertainty of the jet energy scale (JES).
Typical JES uncertainties at CDF and \dzero are about 2\%, which could lead to an uncertainty of about 2 GeV in the measured \mtop.
One way to get around this is to use kinematic variables that are insensitive to the JES, such as lepton \pt.
Unfortunately in these cases the sensitivity of the measurements to \mtop is also reduced, leading to a much larger statistical uncertainty in the measured \mtop~\cite{CDFPt}.
A different approach, the so-called {\sl in situ} JES calibration, has been developed for the lepton+jets and all hadronic channels.
Here a global factor on the JES is constrained by using the two jets produced from hadronic $W$ decay and the well-measured \mW value.
The uncertainty on \mtop due to the JES uncertainty can be reduced by a factor of 2 or more by using the {\sl in situ} JES calibration, as reported in recent measurements.

Another challenge in direct top-quark mass measurements comes from the jet-parton assignment.
For example, most measurements with the template method use the reconstructed top-quark and $W$ masses to build the MC templates and compare with data.
In order to be able to reconstruct the top-quark and W one has to know which jet comes from which parton.
In the lepton+jets channel with 4 quarks in the final state, the number of jet-parton permutations that need to be considered is 12.
This number can be as high as 180 in the all hadronic channel.
By requiring one or more jets that are identified as $b$-quark jets, not only the background contribution is reduced, the number of jet-parton permutations is also greatly reduced.
Another approach to reduce the number of possible jet-parton permutations is based on kinematic fitters, which compare the difference between the measured kinematic variables and expected ones with the measurement uncertainties. 
Only the best jet-parton permutation(s) determined by the kinematic fitter is considered.
Finally one can also combine the results from all the jet-parton permutations using the likelihood of each permutation being the correct jet-parton assignment.

The two most recent \dzero results are both based on the ME method.
The first one is obtained in the lepton+jets channel using 3.6 \fb of data~\cite{D0ljets}.
Exactly 4 jets in one event is required, with at least one of the jets being identified as a $b$-quark jet.
The dominate background contribution is $W$+jets production, which is modeled using \alpgen MC generator interfered with \pythia for parton showering and hadronization.
The measurement combines an {\sl in situ} JES calibration with the standard JES derived in studies of $\gamma$+jet and dijet event. 
Compared to previous measurements at \dzero, a major improvement in this new measurement is the significant reduction of the uncertainty associated with the modeling of differences in the calorimeter response to $b$-quark and light-quark jets originating from the introduction of a new flavor-dependent jet energy response correction.
The measurement gives \mtop$=174.9\pm0.8(stat)\pm0.8(JES)\pm1.0(syst)$ GeV.
This is the best top-quark mass measurement at \dzero.
The second result comes from the dilepton channel using 5.4 \fb of data~\cite{D0dilepton}.
Exactly 2 jets in one event is required without explicit $b$-quark jet identification.
The dominant background contribution comes from $Z/\gamma^*$+jets production.
An {\sl in situ} JES calibration is not applied since there is no hadronically decayed $W$.
Thus the dominant uncertainty is coming from JES systematic uncertainties.
The measurement gives \mtop$=174.0\pm1.8(stat)\pm2.4(syst)$ GeV.
This is the best top-quark mass measurement in the dilepton channel in the world.

The two most recent CDF results both use the template method with {\sl in situ} JES calibrations.
The first one is obtained in the all hadronic channel using 5.8 \fb of data~\cite{CDFAllHadronic}.
The event selection uses a Neural Network (NN) trained on \ttbar signal MC.
Exactly 6 to 8 jets in an event are required with at least one of them being identified as $b$-quark jet.
The background is dominated by multijet production, and is modeled using data and $b$-quark jet tag rate obtained from events with exactly 5 jets.
A kinematic fitter is used to reconstruct the top-quark and $W$ masses, which are then used to build the MC templates to compare with data.
The measurement gives \mtop$=172.5\pm1.4(stat)\pm1.0(JES)\pm1.1(syst)$ GeV.
This is the second best measurement at CDF.
The second new CDF result comes from the MET+jets channel using 5.7 \fb of data~\cite{CDFAllMETJet}.
Events are collected by a multijet trigger, and are required to have 4 to 6 jets, 0 lepton, and large \ptmiss.
NN and $b$-quark jet identification are used to suppress background contribution.
It is found in MC studies that most of the signal events come from the \ttbar lepton+jets decay channel, in which the lepton escapes the detection or is a $\tau$ lepton which decays hadronically.
The measurement gives \mtop$=172.3\pm1.8(stat)\pm1.5(JES)\pm1.0(syst)$ GeV.
It provides sensitivity complementary to the other top-quark mass measurements.

The latest combination of the direct top-quark mass measurements from the Tevatron gives the result \mtop$=173.3\pm0.6(stat)\pm0.9(syst)$ GeV~\cite{tevewwg}, corresponding to a relative uncertainty of 0.6\%.
As the uncertainty is dominated by the systematic uncertainty, programs are on-going to understand better the sources of the systematic uncertainties in order to achieve an better accuracy.
This Tevatron combination was performed in July 2010, and did not include any of the above mentioned results nor the updated best CDF result in the lepton+jets channel using the ME method~\cite{CDFljets}.
It is expected that with the full Tevatron RunII data of about 10 \fb, the total uncertainty in the measured \mtop will be below 1.0 GeV.
More direct top-quark mass results from CDF and \dzero can be found at public web pages \cite{MassPages}.

\section{Determination of Top-Quark Mass from Top Pair Production Cross Section\label{xsection}}
Beyond Leading Order (LO) Quantum Chromodynamics (QCD), the mass of the top quark is a parameter depending on the renormalization scheme.
The definition of mass in field theory can be divided into two categories: (i) driven by long-distance behavior, which corresponds to the pole-mass scheme, and (ii) driven by short-distance behavior, which for example is represented by the \MSbar scheme.
In the direct measurements of the top-quark mass, the quantity measured in data is calibrated w.r.t.~assumed top-quark mass values in MC, thus corresponds to the scheme in the MC simulation.
The top quark mass scheme in MC has not been directly connected with the pole or \MSbar scheme, although it has been argued that the top quark mass in MC scheme should be close to the pole mass~\cite{Fleming}.

\dzero has updated the study on determining the top-quark pole mass and \MSbar mass through top pair production cross section ($\sigma_{\ttbar}$)~\cite{MassXSection}.
In this study, a likelihood $L(\mtop)$ is calculated by comparing the measured $\sigma_{\ttbar}$ in the lepton+jets channel using 5.3 \fb of data with next-to-leading-order (NLO) or next-next-to-leading-order (NNLO) calculations:
\begin{equation}
L(\mtop)=\int f_\mathrm{exp}(\sigma|\mtop)[f_\mathrm{scale}(\sigma|\mtop)\otimes f_\mathrm{PDF}(\sigma|\mtop)],
\end{equation}
where $f$'s are probability functions of $\sigma_{\ttbar}$ and \mtop, determined from \dzero measured $\sigma_{\ttbar}$ with weak dependence on \mtop in MC through acceptance and detection efficiency effects ($f_\mathrm{exp}$), or from theoretical calculations using top quark pole mass or \MSbar mass with renormalization and PDF uncertainties ($f_\mathrm{scale}\otimes f_\mathrm{PDF}$).
The \mtop extracted from $L(\mtop)$, under the assumption \mtop in MC equal to the top quark pole mass, is found to agree with the average value of \mtop from the Tevatron combination~\cite{tevewwg}, while the \mtop extracted assuming \mtop in MC equal to the top quark \MSbar mass is found to be different from the average Tevatron value.
The uncertainty in such extracted \mtop, which is dominated by systematic uncertainties in the measured $\sigma_{\ttbar}$, is quite large to make a more quantitative statement.

\section{Measurements of Top-Antitop Quark Mass Difference\label{difference}}
\dzero published the first measurement of the top-antitop mass difference using a ME method on 1 \fb of data, and found $\Delta M=3.8\pm3.4(stat)\pm1.2(syst)$ GeV~\cite{D0MassDifference1}.
The top and antitop quark masses ($M_t$ and $M_{\bar{t}}$) are measured independently.
Recently, CDF has also contributed a measurement using a template method on 5.6 \fb of data, and found $\Delta M=-3.3\pm1.4(stat)\pm1.0(syst)$ GeV~\cite{CDFMassDifference}.
The result deviates from the expectation of CPT invariance, $\Delta M=0$ GeV, at about 2$\sigma$ level.
In the CDF measurement, $M=(M_t+M_{\bar{t}})/2$ is constrained to be 172.5 GeV.
\dzero has updated the measurement using 3.6 \fb of data, and found $\Delta M=0.9\pm1.8(stat)\pm0.9(syst)$ GeV~\cite{D0MassDifference}.
The result is consistent with the expectation of CPT invariance.
The top-antitop quark mass difference measurements are dominated by statistical uncertainty.
The uncertainty is expected to be improved with the full Tevatron RunII data.


\section*{References}
\begin{thebibliography}{99}
\bibitem{topdiscovery} S. Abachi \etal (D0 Collaboration), \Journal{\PRL}{74}{2632}{1995}; F. Abe \etal (CDF Collaboration), \Journal{\PRL}{74}{2626}{1995}.

\bibitem{lepewwg} The ALEPH, CDF, D0, DELPHI, L3, OPAL, SLC Collaborations, the LEP Electroweak Working Group, the Tevatron Electroweak Working Group, and the SLD electroweak and heavy flavour groups, arXiv:hep-ex/1012.2367v2, (2011); LEP Electroweak Working Group, \url{http://lepewwg.web.cern.ch/LEPEWWG/}.

\bibitem{tevewwg} The Tevatron Electroweak Working Group, arXiv:hep-ex/1007.3178v1, (2010); Tevatron Electroweak Working Group, \url{http://tevewwg.fnal.gov}.

\bibitem{Silvia} See Silvia Amerio's contribution in the same conference for details.

\bibitem{CDFPt} CDF Collaboration, CDF public note 9881 (2009).

\bibitem{D0ljets} D0 Collaboration, accepted by {\PRD}, arxiv:1105.6287 (2011).

\bibitem{D0dilepton} D0 Collaboration, accepted by {\PRL}, arxiv:1105.0320 (2011).
\bibitem{CDFAllHadronic} CDF Collaboration, CDF public note 10456 (2011).

\bibitem{CDFAllMETJet} CDF Collaboration, CDF public note 10433 (2011).

\bibitem{CDFljets} CDF Collaboration, \Journal{\PRL}{105}{252001}{2010}.

\bibitem{MassPages} D0 Collaboration, \url{http://www-d0.fnal.gov/Run2Physics/top/top_public_web_pages/top_public.html#mass}; CDF Collaboration, \url{http://www-cdf.fnal.gov/physics/new/top/public_mass.html}.

\bibitem{Fleming} S.~Fleming, A.~H.~Hoang, S.~Mantry and I.~W.~Stewart, \Journal{\PRD}{77}{074010}{2008}.

\bibitem{MassXSection} D0 Collaboration, accepted by {\PLB}, arXiv:1104.2887 (2011).

\bibitem{D0MassDifference1} D0 Collaboration, \Journal{\PRL}{103}{132001}{2009}.
\bibitem{CDFMassDifference} CDF Collaboration, \Journal{\PRL}{106}{152001}{2011}.
\bibitem{D0MassDifference} D0 Collaboration, submitted to {\PRD}, arxiv:1106.2063 (2011).

\end{thebibliography}
\end{document}